# On the importance of AI research beyond disciplines

Virginia Dignum[1], Donal Casey, Teresa Cerratto-Pargman, Frank Dignum, Valentina Fantasia, Bodil Formark, Björn Hammarfelt, Gunnar Holmberg, André Holzapfel, Stefan Larsson, Amanda Lagerkvist, Nicolette Lakemond, Helena Lindgren, Fabian Lorig, Ana Marusic, Lina Rahm, Yulia Razmetaeva, Sverker Sikström, Kıvanç Tatar, Jason Tucker

*Abstract*

*As the impact of AI on various scientific fields is increasing, it is crucial to embrace interdisciplinary knowledge to understand the impact of technology on society. The goal is to foster a research environment beyond disciplines that values diversity and creates, critiques and develops new conceptual and theoretical frameworks. Even though research beyond disciplines is essential for understanding complex societal issues and creating positive impact it is notoriously difficult to evaluate and is often not recognized by current academic career progression. The motivation for this paper is to engage in broad discussion across disciplines and identify guiding principles fir AI research beyond disciplines in a structured and inclusive way, revealing new perspectives and contributing to societal and human wellbeing and sustainability.*

The results of Artificial Intelligence (AI) research are increasingly becoming a part of many scientific fields, affecting the way of conducting research and generating knowledge. AI brings new tools and methodologies, new ways of thinking and new ontologies to the research community. Hence, it is also becoming increasingly important to establish a bidirectional relationship between the disciplines developing, implementing, and promoting new AI technologies and those that are adapting to AI technological advances. Instead of *pushing* new AI developments that potentially affect other disciplines and that might not be aligned with their individual research paradigms and methodologies, it is crucial that disciplines affected by AI advances have the ability to feed back their demands, requirements, and concerns, to actively and efficiently govern AI development.

Scientific knowledge in the humanities and social sciences has so far been predominantly driven by the conceptualisation of intelligence as a human property. However, current societal challenges posed by artificial intelligence (AI) technologies require new research perspectives and novel, collaborative ways to look at the world (D'Ignazio & Klein, 2019). AI systems cannot be studied as humans, and therefore demand new scientific methodologies and theories in general and from the humanities and social sciences in particular. Science needs to fundamentally adapt its focus, research questions, data and knowledge, to include implications related to both artificial systems and natural intelligence and behaviour. At the same time, the main communication channels in society have been shifting, and this shift is initiating a need for evolution of academic knowledge dissemination, to ensure the societal relevance of scientific knowledge.

---

[1] The first author is the initiator of the project. All other authors are listed alphabetically and contributed equally to the paper.

Meeting the societal, cultural, and scientific challenges of AI requires courageous and collaborative research that cultivates and engages with a diversity of perspectives. Such an approach to research should advance new methods to analyse, design, develop, evaluate, critique and assess the human and societal impact of technology. Having the openness and respect to learn the contributions that perspectives of other disciplines, and being humble enough to reflect upon our disciplinary limitations, can bring and trying to combine them into a new approach for current problems is key to success. Going beyond disciplinary research on AI and autonomous systems is a much-needed step towards understanding and addressing their impact on humanity and society (Dignum, 2020). Research beyond disciplines should primarily be assessed by its capacity to generate positive impact and justice for both the individuals and the society, as well as its ability to push debates beyond the status quo and to propose alternatives on how to address societal consequences.

Understanding and shaping the impact of AI on the individuals and the society also calls for novel interdisciplinary knowledge in the humanities and social sciences to address fundamental, conceptual, practical and creative challenges about the interdependencies between technology and the ways humans relate and interact, form social institutions, shape decision making and rebuild their environment. Such endeavours will generate cutting-edge research that harnesses expertise, build competence, foster knowledge development beyond disciplinary boundaries, and contribute to the development of innovative humane technologies and their transformative interaction with society. Going beyond established research disciplines is not so much a goal in itself, but the central means to study and analyze different facets of a given issue, thus developing new ways to understand the world and bringing about positive change in both the research community and society.

An important component of our mission as researchers in the field of AI and its societal impact is to foster a research environment beyond disciplines that will encourage the development of novel, creative, pluralistic conceptual and theoretical frameworks that value curiosity, diversity, and care. Our research encompasses, merges and extends various disciplines, application areas and stakeholders. As researchers in AI coming from many different disciplinary backgrounds, we therefore seek to create an environment in which research beyond disciplinary boundaries is enabled, recognised, and valued. We therefore seek to establish 'guidelines' to address and structure new opportunities for collaborative research between disciplines that often have not worked together before.

Research beyond disciplines is much-needed to move towards new knowledge, encourage novel understanding of complex societal issues, and enable the grounding or framing of research on complex 'real world' questions as well as on sound scientific theories and models. The diversity of societal, environmental and humane expressions cannot be engaged from a single cultural interpretative framework (Ihde, 1993, 114-115). Research beyond disciplines is not only rewarding academically but is also impactful for people, society, culture, and the planet. Studying technologies should be conceived as a post-disciplinary practice that enables cross-cutting lines of inquiry, embraces pluralism, and addresses real-world problems (Silvio Waisbord, 2019). Such practice strongly resonates with our intention to move beyond the disciplines.

Research beyond existing disciplinary boundaries is however hard to realise. It is difficult to evaluate using conventional methods for assessing and measuring research. Disciplinary evaluation of research is based on the techniques, models, and methods that define that discipline. In research beyond disciplines, each project must create its own methodology based on the practices of all disciplines involved. That takes time and needs to be done very carefully with understanding of, and respect for the traditions of each discipline. It especially requires open minds that are not afraid to explore new territory that may not traditionally be valued by some of the disciplines. Excellence in research beyond disciplines is evident from its contributions to knowledge creation and societal impact, rather than solely fitting existing scientific methods. Nevertheless, currently, academic career progression is closely tied to the contribution to existing disciplines, through research outputs, funding, teaching and other areas of professional practice. This needs to change in order to recognise the contributions of research beyond disciplines. A core motivation for doing research beyond disciplines is cultivating our ability to engage in societal challenges and discussions for which we want to build and contribute knowledge, and to do this in a structured, participatory and inclusive way, challenging our own personal boundaries. Researching beyond disciplines can reveal new perspectives on timely issues, by focusing on the same topic through a diversity of lenses.

Collaboration rather than competition lies at the core of successful and meaningful research beyond disciplines. This stands in stark contrast with current systems of research assessment and evaluation that valorise, prioritise and reward competition and individual excellence. The emphasis on competition is not compatible with a system that values contribution to society, which is implicated, effected and entangled in the work we do as researchers. In line with the EU Open Science Policy[2], we see contribution to the society as a fundamental part of open and responsible science, and call for its consideration when evaluating research excellence.

We call for novel evaluation mechanisms that give importance to exploration of new research grounds and to the impact of research on society, and that are inclusive, recognizing the power relations underlying many collaborations. When going beyond disciplines, the generation of new research questions is more important than testing the initial hypotheses to determine whether they confirm the explanatory theory. Excellence in this context is not standing out as an individual (person or group) but the capacity to bring together different voices, pushing forward through unexplored fields and learning from failure.

Research beyond disciplines is imperative to address complex societal questions that arise at the nexus between the living and technology, focusing on the same problem for which we want to build knowledge from different viewpoints provides new perspectives and contributes to all disciplines. We therefore propose the following guiding principles for AI research beyond disciplines:

- Recognise the contribution of research beyond disciplines to knowledge creation is at least as important as that of single disciplines, and it has the added value of discovering and addressing complex issues that 'fall in between' the lines.
- Scrutinise the use single-disciplinary research, rather than approaches outside existing disciplinary boxes. The reviewing processes for research work on AI and autonomous systems must standardly include the question "justify why only one discipline was applied?"
- Measure the quality of research beyond disciplines by its capacity to shape the conversation, engage stakeholders, and contribute to a process of change, and not just through publication statistics and individual rankings.

- Evaluate the novelty and transformative potential of ideas, theories, methodologies, solutions, and the approaches they offer and how these can stimulate public discussion.
- Avoid pressuring researchers to determine what discipline they belong to or what discipline their research belongs to and then assessing whether it fits the stated framework.
- Value exploration and approach disciplinary tensions as a resource for reflexivity, creativity, innovation and knowledge development, which address "common matters of concerns" rather than disciplinary matters or "matters of facts" (Latour, 2004).
- Embrace a broad framework of critique and just development that goes beyond technological solutionism.

We invite all researchers working on the design, development, analysis, critique and evaluation of AI and its societal impact, to join us in discussing and implementing these guidelines.

**References**

D'Ignazio, C. & Klein, L. (2019). Conclusion: 'Now let's multiply' (pp. 203-214) in *Data Feminism*. Cambridge: MIT Press.

Dignum, V. 2020. AI is multidisciplinary. *AI Matters* 5, 4 (December 2019), 18–21. https://doi.org/10.1145/3375637.3375644

Ihde, D. (1993). *Postphenomenology: Essays in the postmodern context*. Evanston, Il: Northwestern University Press.

Latour, B. (2004). Why has critique run out of steam? From matters of fact to matters of concern. *Critical inquiry*, *30*(2), 225-248.

Waisbord, S. (2019) *Communication: A Post-discipline*, Oxford: Polity.

**Authors' information:**

- (Contact author) Virginia Dignum, Department of Computing Science, Umeå University, Sweden, virginia@cs.umu.se
- Ingar Brinck, Department of Philosophy, Lund University, Sweden
- Donal Casey, Department of Business Studies, Uppsala University, Sweden
- Teresa Cerratto-Pargman, Dept. of Computer and Systems Sciences, Stockholm University, Sweden
- Frank Dignum, Department of Computing Science, Umeå University, Sweden
- Valentina Fantasia, Department of Philosophy, Lund University, Sweden
- Bodil Formark, Umeå University, Sweden
- Björn Hammarfelt, Swedish School of Library and Information Science, University of Borås
- Gunnar Holmberg, Department of Management and Engineering, Linköping University, Sweden
- André Holzapfel, Division of Media Technology and Interaction Design, KTH, Sweden
- Stefan Larsson, Department of Technology and Society, Lund University, Sweden
- Amanda Lagerkvist, Department of Informatics and Media, Uppsala University, Sweden
- Nicolette Lakemond, Department of Management and Engineering, Linköping University, Sweden
- Helena Lindgren, Department of Computing Science, Umeå University, Sweden

- Fabian Lorig, Department of Computer Science and Media Technology, Malmö University, Sweden
- Ana Marusic, University of Split School of Medicine, Split, Croatia
- Lina Rahm, Department of Philosophy and History, KTH, Sweden
- Yulia Razmetaeva, Department of Law, Uppsala University, Sweden
- Sverker Sikström, Department of Psychology, Lund University, Sweden
- Kıvanç Tatar, Chalmers University of Technology, Sweden
- Jason Tucker, Department of Global Political Studies, Malmö University, Sweden